# Geometry Effects on Switching Currents in Superconducting Ultra Thin Films

A. Leo*, A. Nigro, S. Pace
Physics Department
University of Salerno
Fisciano (SA), Italy
*antonio.leo@fisica.unisa.it

A. Guarino, N. Martucciello, G. Grimaldi°
SPIN Institute
CNR
Salerno, Italy
°gaia.grimaldi@spin.cnr.it

V. La Ferrara, E. Bobeico
CR Portici
ENEA
Portici (NA), Italy

J.C. Villégier
CEA
Grenoble
Grenoble, France

*Abstract*— Vortex dynamics is strongly connected with the mechanisms responsible for the photon detection of superconducting devices. Indeed, the local suppression of superconductivity by photon absorption may trigger vortex nucleation and motion effects, which can make the superconducting state unstable. In addition, scaling down the thickness of the superconducting films and/or the width of the bridge geometry can strongly influence the transport properties of superconducting films, e.g. affecting its critical current as well as its switching current into the normal state. Understanding such instability can boost the performances of those superconducting devices based on nanowire geometries. We present an experimental study on the resistive switching in NbN and NbTiN ultra-thin films with a thickness of few nanometers. Despite both films were patterned with the same microbridge geometry, the two superconducting materials show different behaviors at very low applied magnetic fields. A comparison with other low temperature superconducting materials outlines the influence of geometry effects on the superconducting transport properties of these materials particularly useful for devices applications.

*Keywords— critical currents; flux pinning; superconducting detectors; superconducting thin films*

## I. Introduction

Many superconducting detectors require high-performing thin films able to carry high critical currents [1] and also to get a fast transition to the normal state [2]. A superconducting detector is usually made up by a single, or an array of long nanowires of ultrathin films. For example, the operation principle of a single photon detector is based on the transition from the superconducting state to the normal state of a small part of the device hit by the photon. Among the low superconducting materials commonly used for this kind of application [1-5], we picked up the mostly used, i.e. NbN and NbTiN ultra-thin films [6,7], in order to study geometry effects on the switch response of the material [8,9]. In particular, we focused on the abrupt transition to the normal state marked by a critical voltage $V^*$ and an instability current $I^*$ which can be measured in the current-voltage characteristics of the superconducting samples patterned by standard UV lithography in a microbridge geometry. By applying an increasing bias current, the microbridge can be suddenly driven from the superconducting state up to the normal state [10]. Such an abrupt transition can mimic the transition due to a photon detection. Indeed, the hot-spot formation is the mechanism to which the photon detection is usually ascribed, and this mechanism key-point is that the superconducting state becomes unstable, suddenly switching to the normal resistive state. Therefore, the study of this instability can provide a useful framework in which the critical parameters $I^*$ and $V^*$ can be analyzed as a function of the operational temperature, of the bias current and of low magnetic fields comparable to those induced by the bias current [10,11].

## II. Experimental Results

First, high quality NbN and NbTiN ultra-thin films with a thickness of 5 nm, epitaxially grown on sapphire substrate [6], have been used to obtain several microbridges by standard UV lithography. The microbridges are 1 mm long and 50 μm wide. All samples were characterized by a pulsed current 4-probe technique in a CFM16T cryogen-free system by Cryogenic, Ltd. equipped with a 16 T magnet. Current-voltage (I-V) characteristics have been acquired at fixed temperatures of 1.6 K and 4.2 K, and as a function of the applied magnetic field in a range around 0 T, namely -10 mT to 500 mT. An accurate evaluation of the residual trapped magnetic field has been achieved by the analog signal from a local Hall sensor. Resistance-vs-temperature curves were acquired in a fixed applied magnetic field up to values of 16 T.

In Fig. 1a) and 1b) the field-vs-temperature phase diagrams of two samples, namely NNA_1 and NTNB_1, are reported in the whole range of applied magnetic field values, for the two superconducting materials NbN and NbTiN, respectively. The field is applied perpendicular to the film surface. In each phase diagram three curves are displayed: the $H_{c2}$ line for the upper critical field values evaluated at the 90% of the normal state resistance, the $T_C$ line for the determination of the critical temperature parameter based on the 50% of the normal state resistance criterion, and the $H_{irr}$ line determined at the 10% of the normal state resistance.

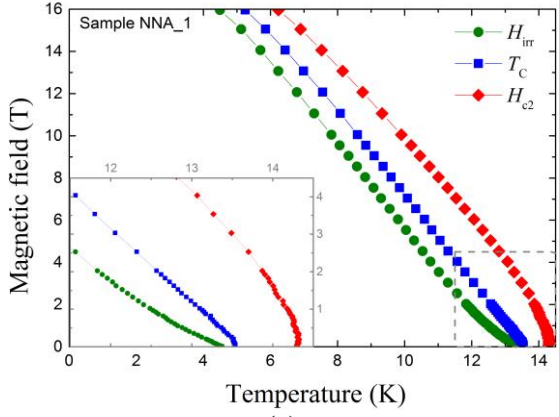

(a)

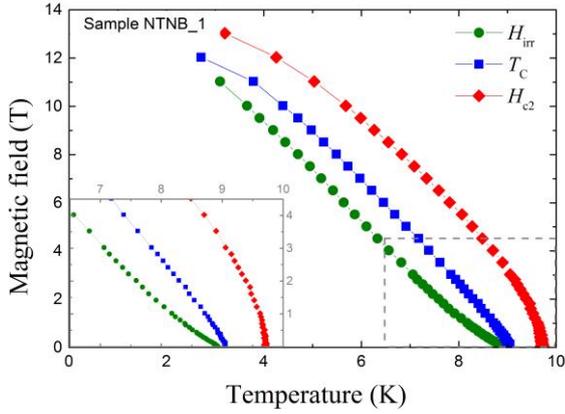

(b)

**Fig. 1:** Magnetic field-temperature phase diagrams of two investigated samples: (a) NNA_1 and (b) NTNB_1 in the full field range. The insets show an enlargement view of the dotted confined region close to the critical temperature.

However, the present study is focused on a small field range around 0 T. In this field range I-V measurements have been performed in order to push the system out of the equilibrium conditions by a proper bias operational mode [10]. Typical I-V curves recorded at a fixed temperature are shown in Fig. 2a) and 2b) for the two samples NNA_1 and NTNA_1, respectively.

The critical values of the current and voltage instability point are marked in the Fig. 2 as ($I^*$, $V^*$). This instability point corresponds to the switching from the superconducting state up to the normal state for each microbridge under investigation. The symmetry of the I-V characteristics, as well as the absence of thermal hysteresis is well established [10].

The experimental result reported in Fig. 3 concerns the critical vortex velocity $v^*$ as a function of the local actual magnetic field applied perpendicularly to the film surface.

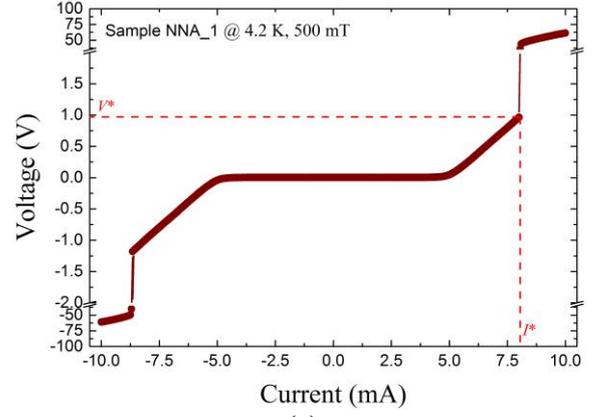

(a)

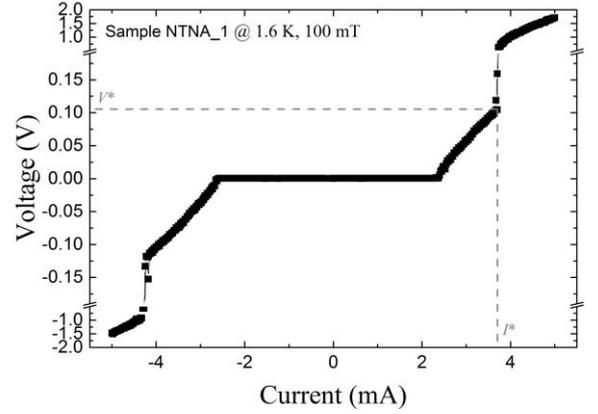

(b)

**Fig. 2:** I-V characteristics measured on NbN (a) and NbTiN (b) samples in the full bias current range up to the normal resistance state. The critical parameters ($I^*$, $V^*$) are marked.

This critical parameter is a measure of the critical flux flow voltage $V^* = v^* \cdot H \cdot l$, where $H$ is the applied magnetic field and $l$ is the microbridge length.

### III. DISCUSSION AND CONCLUSIONS

Despite the similar physical parameters of the two investigated materials NbN and NbTiN, which have already been established to be suitable for single photon detection applications [12,13], a comparison of the critical parameters here investigated outlines that they have a strongly different behavior in the presence of very small magnetic fields. In particular, the critical vortex velocity shows an opposite behavior around 0T, regardless of the same geometrical parameters employed (see Fig. 3). Indeed the field dependence of $v^*$ observed for the NbN, i.e. NNA_1 sample, can be fairly explained by the low field behavior found in other low $T_C$ superconducting materials [14-16] and in mesoscopic superconductors [8]. In fact, the mesoscopic regime is achieved also for these materials, due to the chosen geometry: ultra-thin film with a wide microbridge. On the contrary, the opposite dependence found for NbTiN, i.e. NTNB_1, can be ascribed to the different intrinsic pinning role developed by Ti addition [14,15]. In conclusions, although the present study can confirm that the superconducting transport properties of these materials are not affected by the ultra-thin geometry [2,6], a remarkable geometry effect is emphasized by making the superconducting microbridge thinner and not narrower.

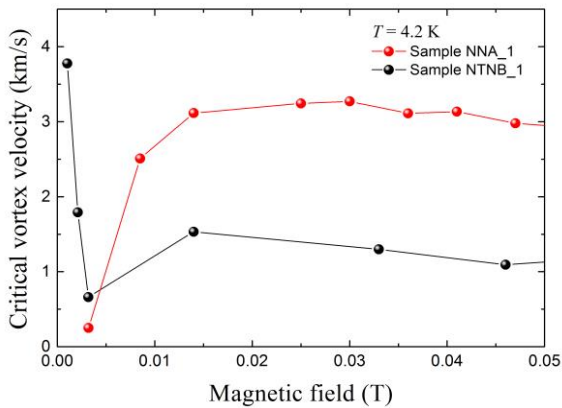

**Fig. 3:** A comparison of the magnetic field dependence of the critical parameter $v^*$ between the two investigated materials.

ACKNOWLEDGMENT

The students L. Genovese and G. Luongo have been involved during samples resistivity measurements.